\documentclass[10pt,a4paper,onecolumn]{article}
\usepackage{graphicx}
\usepackage{xcolor}
\usepackage{authblk}
\usepackage{titlesec}
\usepackage{hyperref}
\usepackage{amssymb,amsmath}
\usepackage{float}
\usepackage{lscape}
\usepackage[backend=biber,style=chicago-authordate,url=false,doi=false,isbn=false]{biblatex}
\bibliography{paper.bib}

\usepackage[top=3.5cm, bottom=3cm, right=1.5cm, left=1.0cm]{geometry}

\title{Cost-Effective Big Data Orchestration Using Dagster: A Multi-Platform Approach}
\author[1]{Hernan Picatto}
\author[1, 2]{Georg Heiler}
\author[1, 2, 3]{Peter Klimek}
\affil[1]{Supply Chain Intelligence Institute Austria}
\affil[2]{Complexity Science Hub}
\affil[3]{Section for Science of Complex Systems, Center for Medical}
\date{\today}

\begin{document}
\maketitle

% Adding the previously marginpar content into the main body
\noindent
\textbf{DOI:} \href{https://doi.org/DOI unavailable}{DOI unavailable}\\
\textbf{Software:} \href{N/A}{Review}, \href{https://github.com/ascii-supply-networks/ascii-hydra}{Repository}, \href{DOI unavailable}{Archive}\\
\textbf{Editor:} Pending Editor\\
\textbf{Reviewers:} Non review version
\textbf{Submitted:} N/A\\
\textbf{Published:} N/A\\
\textbf{License:} Authors of papers retain copyright and release the work under a Creative Commons Attribution 4.0 International License (\href{http://creativecommons.org/licenses/by/4.0/}{CC BY 4.0}).

\hypertarget{summary}{%
\section{Summary}\label{summary}}

The rapid advancement of big data technologies has underscored the need for robust and efficient data processing solutions. 
Traditional Spark-based Platform-as-a-Service (PaaS) solutions, such as Databricks (DBR) and Amazon Web Services Elastic MapReduce (EMR), provide powerful analytics capabilities but often result in high operational costs and vendor lock-in issues. 
These platforms, while user-friendly, can lead to significant inefficiencies due to their cost structures and lack of transparent pricing (Zaharia et al., 2016).

This paper introduces a cost-effective and flexible orchestration framework using Dagster (Dagster, 2018). 
Our solution aims to reduce dependency on any single PaaS provider by integrating various Spark execution environments. 
We demonstrate how Dagster's orchestration capabilities can enhance data processing efficiency, enforce best coding practices, and significantly reduce operational costs. 
In our implementation, we achieved a 12\% performance improvement over EMR and a 40\% cost reduction compared to DBR, translating to over 300 euros saved per pipeline run.

Our goal is to provide a flexible, developer-controlled computing environment that maintains or improves performance and scalability while mitigating the risks associated with vendor lock-in. 
The proposed framework supports rapid prototyping and testing, which is essential for continuous development and operational efficiency, contributing to a more sustainable model of large data processing.

\hypertarget{statement-of-need}{%
\section{Statement of need}\label{statement-of-need}}

In the domain of large-scale data processing, Platform as a Service (PaaS) solutions based on the Spark framework, such as DBR, have demonstrated notable advantages in terms of user-friendliness and robust analytical capabilities. 
However, these benefits are not without substantial drawbacks, including high costs associated with vendor lock-in and a lack of transparent pricing, which can escalate rapidly and unpredictably. 
The convenience of such platforms can result in inefficient deployments, where the ease of adding resources does not correlate with need or cost-efficiency. 
This can diminish developer productivity, hinder good development practices, and increase operational expenses.

The proposed solution employs Dagster's orchestration capabilities to facilitate a robust integration of diverse Spark environments, thereby reducing reliance on a single PaaS provider. 
This approach mitigates the potential for lock-in risks, enhances cost efficiency, and incorporates coding standards and best practices into the data science pipeline. 
This transition not only enhances productivity by enabling expeditious prototyping and testing on smaller data sets but also markedly reduces costs by optimising resource utilisation across disparate platforms without compromising performance

This approach is of particular importance for organisations seeking to maintain agility and scalability in their data operations without incurring excessive costs. 
By decentralising the data processing tools and focusing on enhancing orchestration and flexibility, organisations can achieve more sustainable growth and innovation in the domain of big data.

This structured approach ensures consistency across the various stages of development and facilitates the verification and replication of results, which are critical elements in scientific research. 
By leveraging a dedicated orchestrator like Dagster, which emphasises containerisation and robust orchestration capabilities, researchers can create workflows that are not only efficient but also transparent and replicable. 
%This is crucial in maintaining the integrity of the scientific process, allowing for validation and extension by other researchers. 
Such systems foster a collaborative scientific environment where methodologies are as open and accessible as the findings they produce.

\hypertarget{relevance}{%
\section{Relevance}\label{relevance}}

The proposed architectural framework, which employs Dagster, facilitates the reproducibility of pipelines by centralising the management of metadata and standardising the orchestration of data processing tasks across diverse operational environments. 
By abstracting the complexity of the underlying infrastructure and providing tools for seamless integration and deployment, the architecture particularly facilitates the replication of scientific experiments under identical conditions.

Notwithstanding the mounting interest in data pipelines, authors such as Mathew et al. (2024) concentrate on the optimisation of big data processing through sophisticated scheduling techniques that minimise energy consumption and latency. 
While their work also aims to optimise resource utilisation in data centres, its core emphasis is on the algorithmic enhancement of scheduling mechanisms, rather than on orchestration across different PaaS solutions or on the promotion of coding practices within data pipelines. 
In their 2021 paper, Daw et al. examine the creation of a framework for automated scaling of resources in cloud environments. 
Their work focuses on aspects of resource allocation based on predictive analytics, with the goal of optimising operational costs and performance. 
In contrast to the work presented here, these approaches do not address the integration of multiple cloud platforms or the orchestration of data processing tasks using open tools like Dagster.

\hypertarget{architecture-model}{%
\section{Architecture Model}\label{architecture-model}}

The framework employs Dagster, an open-source data orchestrator designed for the construction, operation, and observation of data pipelines. 
The decision was taken to run Dagster on Docker in order to maintain greater control over the orchestration environment. 
This configuration enhances the reproducibility and scalability of the system, ensuring alignment with the optimisations made in terms of cost and performance. 
This has the potential to reduce resource consumption by as much as 43\%, see the findings of Heiler \& Picatto (2024) for more details.

More concrete, our objective was to implement a cloud-based management system capable of delivering essential functionalities such as:

\begin{itemize}
\item
  Dynamic resource deployment on the selected platform with automatic scaling;
\item
  Virtual machine and network configuration management;
\item
  Comprehensive deployment and execution monitoring.
\end{itemize}

In order to attain these capabilities, a number of modifications were required.

\begin{figure}[H]
\centering
\includegraphics[width=\textwidth]{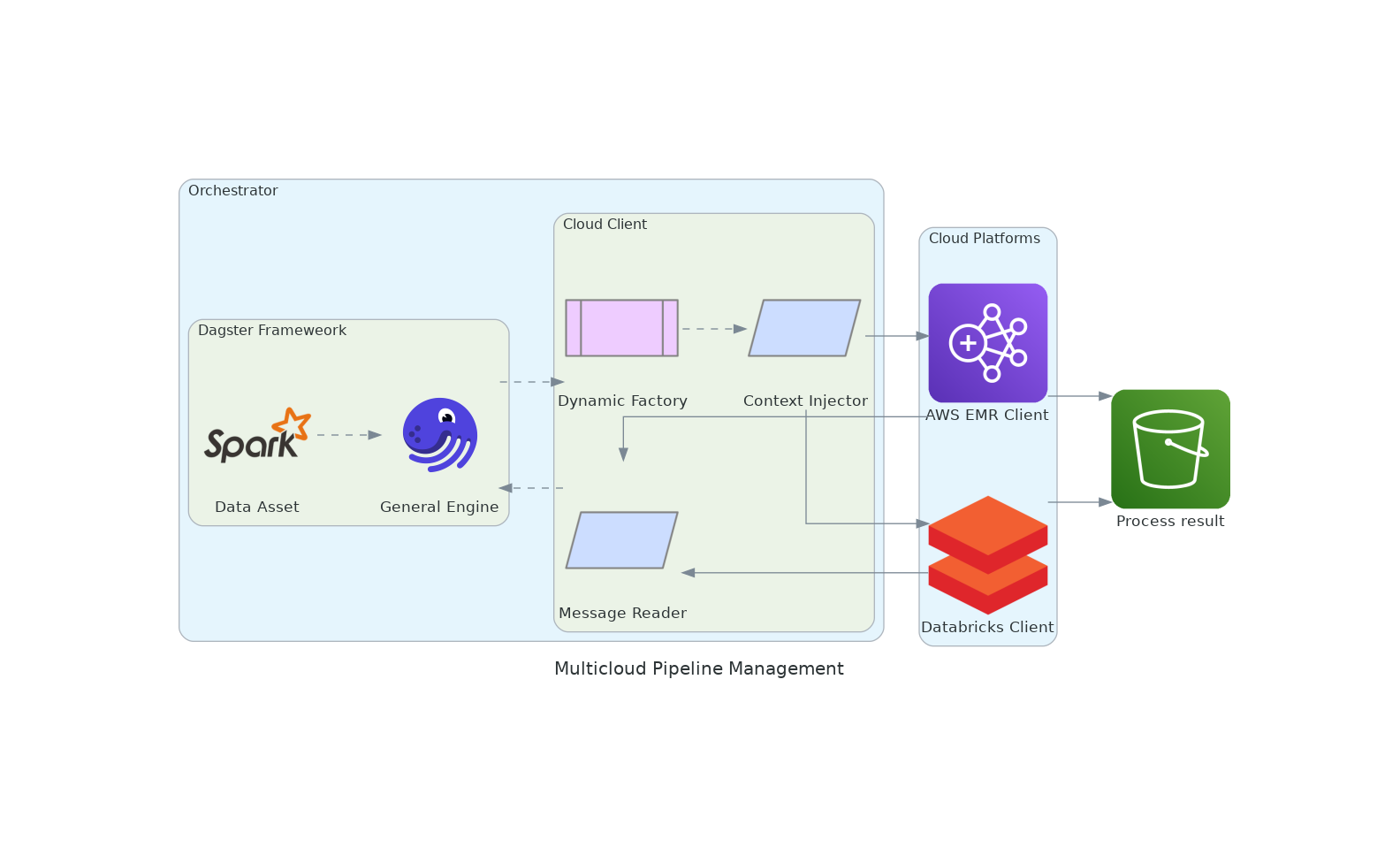}
\caption{Diagram orchestrator behavior.\label{fig:diagram}}
\end{figure}

The core components of our framework include, as illustrated in figure \ref{fig:diagram}:

\begin{enumerate}
\def\labelenumi{\arabic{enumi}.}
\item
  \textbf{Dagster Context Injector}: It oversees the management of general and job-specific configurations, including environmental variables, partitioning, and tagging, which are vital for effective resource management and task segmentation.
\item
  \textbf{Message Reader Improvements}: Optimised to support telemetry functionalities, the system is capable of capturing and processing messages for real-time monitoring and robust debugging, which is particularly useful for EMR.
\item
  \textbf{Cloud Client Innovations}: Introduces a generic cloud client for managing Dagster clients on different platforms, ensuring reliable integration with AWS services and secure environment customisation.
\item
  \textbf{Automation and Integration}: Integrates job definition upload processes with the Databricks REST API and Boto3 clients, automating job setup and environment bootstrapping.
\item
  \textbf{Dynamic Factory for Cloud Client Management}: Detects and designates appropriate execution environments, adapting to changes in processing requirements or processing requirements or platform preferences.
\end{enumerate}

Our strategic enhancements ensure an easy-to-use interface that shields end users from the complexities of directly managing cloud resources. 
This approach significantly reduces the overhead associated with traditional cloud deployments, allowing organisations to focus on strategic tasks.

To ensure production reliability and replicability, this implementation was dockerised. 
Dockerisation enabled a consistent and controlled environment across different development and production stages, minimising inconsistencies and potential problems associated with environment-specific configurations.

\hypertarget{example-use-case-common-crawl-data-processing-and-graph-reconstruction}{%
\section{Example Use Case: Mining web-based interfirm networks from Common Crawl}\label{example-use-case-common-crawl-data-processing-and-graph-reconstruction}}

We illustrate the capabilities of our framework within the task of constructing a web-based mapping of company ecosystems, see also Kinne et al, 2020.
More concrete, we aim to infer different types of relationships between companies by analysing both textual information and hyperlinks found on company websites, which often contain information on how these companies collaborate in innovation activities. 
This mapping involves a detailed extraction process where we identify and pre-process websites from a predefined list of companies (seed nodes), extract edges representing hyperlinks between these nodes, construct a detailed graph and then aggregate this graph to a domain level. 
Each step is critical to understanding the relationships and interactions between companies that are often embedded in the content and structure of their web pages.
The Common Crawl dataset is a particular promising avenue for this kind of research, as it also contains historic data which allows one to map the dynamics of the company ecosystem.

\hypertarget{datasets}{%
\subsection{Datasets}\label{datasets}}

\begin{itemize}
\item
  Common Crawl CC-MAIN: This dataset contains the WARC (Web ARChive) files, which contain the raw web crawl data, and the WAT, which store the computed metadata.
\item
  Seed Nodes: A subset of URLs identified as starting points for our analysis. These nodes are processed to ensure they are relevant and free of common problems.
\end{itemize}

\hypertarget{pipeline-breakdown}{%
\subsection{Pipeline Breakdown}\label{pipeline-breakdown}}

The task requires the combined extraction of text and graph data to also be able to understand due to which activities companies link with each other. 
This necessitates a customized approach to data extraction rather than relying on existing extractions that assume text-only or graph-only data.
%That's because we want to understand why companies have hyperlinks to each other, and to do that we need the graph with its context, neither text only nor graph only would be sufficient. 
Our pipeline includes four key assets:

\begin{enumerate}
\def\labelenumi{\arabic{enumi}.}
\item
  \textbf{NodesOnly}: Extracts and preprocesses seed node information.
\item
  \textbf{Edges}: Extracts HTML content and hyperlinks from seed node
  URLs.
\item
  \textbf{Graph}: Constructs a hyperlink graph by joining nodes and
  edges.
\item
  \textbf{GraphAggr}: Aggregates the graph to the domain level forhigher-level analysis.
\end{enumerate}

\begin{figure}[H]
\centering
\includegraphics[width=\textwidth]{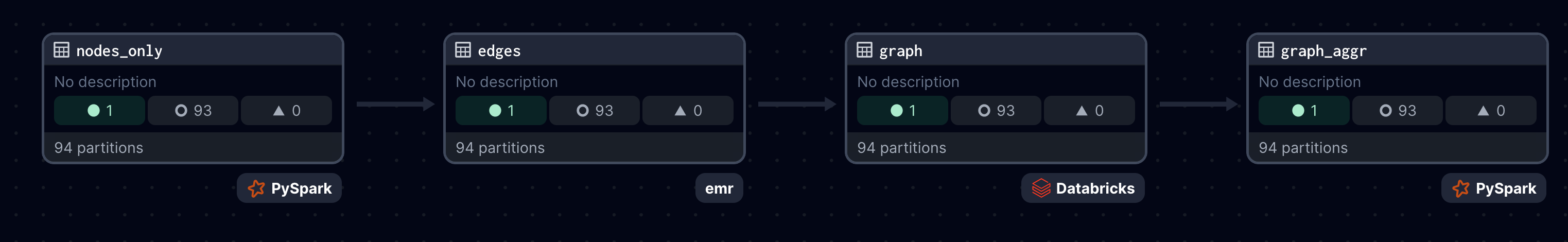}
\caption{Detailed dagster pipeline showcasing how execution environments can be chosen as needed between local, EMR and DBR.
\label{fig:pipleineDagster}}
\end{figure}

Each asset in figure \ref{fig:pipleineDagster} demonstrates the flexibility and efficiency of our orchestration framework, which can adapt to diverse computational needs across different platforms.
Data is partitioned along two primary dimensions: time and domain. 
The temporal partitioning aligns with the Common Crawl\footnote{Common Crawl was accessed between October 2023 and March 2024 from \href{https://registry.opendata.aws/commoncrawl}{Common Crawl}.}  dataset used, facilitating efficient data management and accessibility. 
Conversely, the domain-based partitioning approach is designed to support parallel processing of different research queries. 
This approach allows for the application of different filtering criteria within the data analysis processes, thereby optimising the computational effort and allowing the task to be submitted to the platforms that best suit its needs.

\hypertarget{implementation-challenges}{%
\section{Implementation Challenges}\label{implementation-challenges}}

\begin{figure}[H]
\centering
\includegraphics[width=\textwidth]{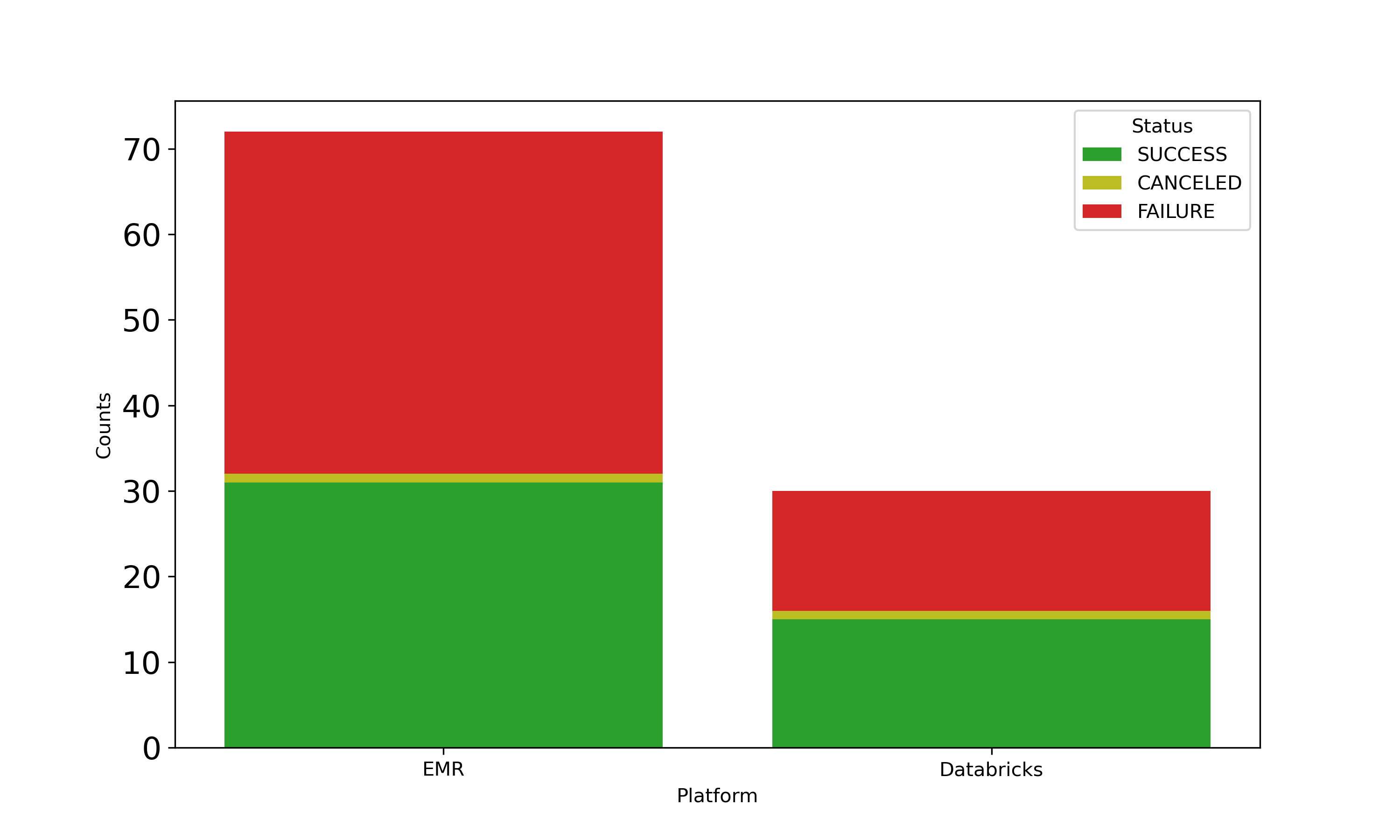}
\caption{Stacked Plot of Trail Runs by
Platform.
\label{fig:stackedTrial}}
\end{figure}

The implementation of new computational platforms, particularly the transition from DBR to AWS EMR, is an example of a complex process that is characterised by a number of challenges and strategic decisions. 
This transition, driven primarily by the potential for cost reduction and enhanced flexibility, reveals significant operational challenges.
Figure \ref{fig:stackedTrial} shows results from runs with sample data, in which run outcomes on both platforms are categorized into three states: success, failure, and cancellation. 
Notably, EMR shows a higher proportion of failures compared to DBR, which is consistent with the ongoing adjustments reflected in the cumulative changes graph. 
This suggests that while EMR provides a cost-effective solution, its operational stability and reliability demand continual oversight and refinement.

Working with EMR (or any not so user friendly PaaS) involves a considerable learning curve and operational investment. 
The initial setup of EMR, despite the foundational familiarity with Spark architectures, proved to be labor-intensive and fraught with technical challenges. 
The necessity for almost double the number of trial runs for EMR as compared to DBR before achieving production stability is indicative of the intricate setup and optimization demands posed by EMR.
This process not only demanded a higher frequency of code revisions but also a deeper engagement with EMR's underlying operational mechanics.

\begin{figure}[H]
\centering
\includegraphics[width=\textwidth]{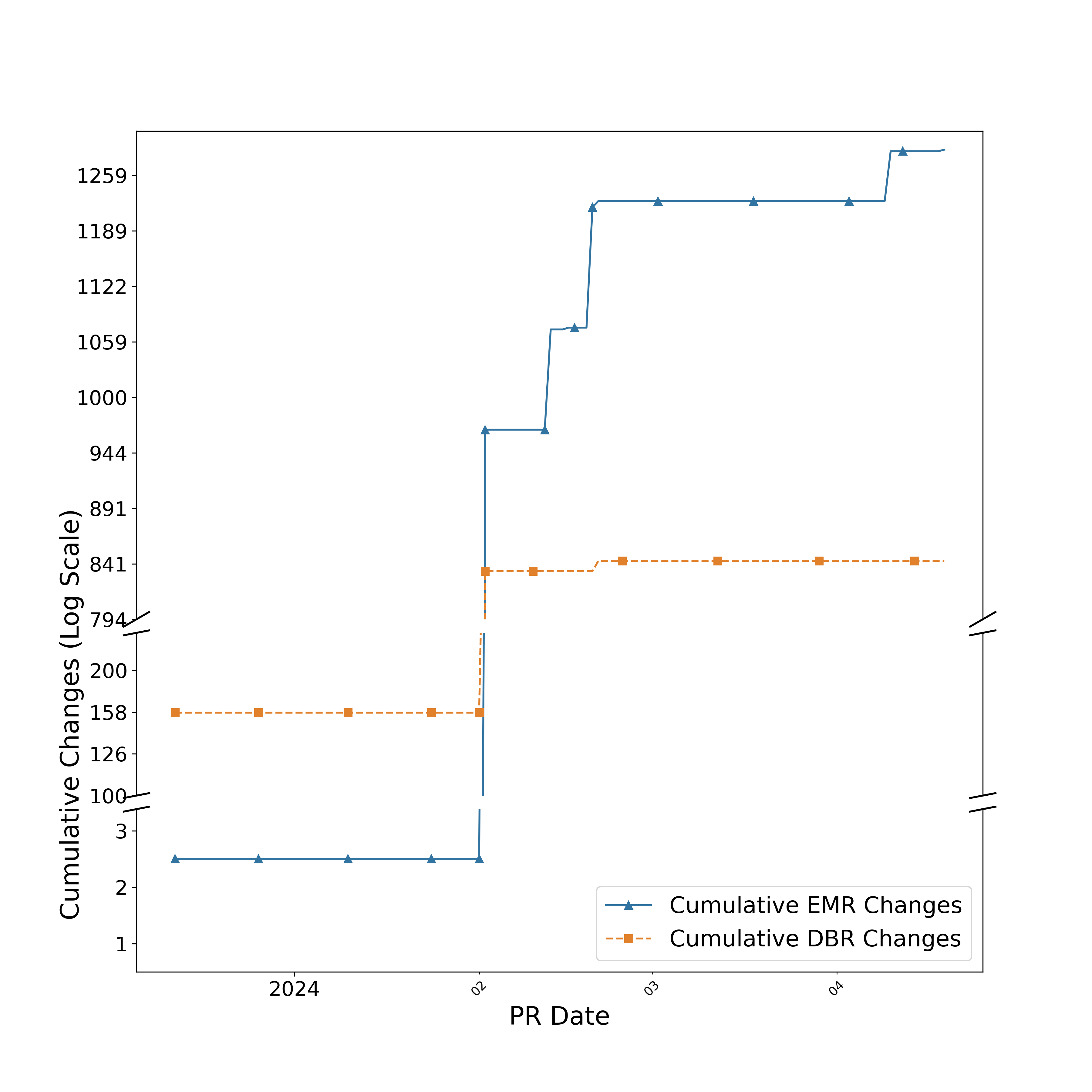}
\caption{Effort Needed for Implementing Each Platform
Client.
\label{fig:linePlatform}}
\end{figure}

The implementation of EMR client exemplifies the considerable learning curve and detailed configuration mastery required to harness its full potential effectively, particularly in comparison to the more user-friendly Databricks platform. 
Through an iterative process, as depicted in the line chart (figure \ref{fig:linePlatform}) showing the cumulative changes over time, our team gradually accrued the technical acumen necessary to fine-tune EMR's environment to meet our rigorous performance and cost-efficiency standards.

This entailed configuring node labelling within YARN in order to guarantee that core nodes, which are more stable, assume responsibility for critical tasks.
Specifically, enabling the \texttt{yarn.node-labels.enabled} parameter and setting \texttt{yarn.node-labels
.am.default-node-label-expression} to \texttt{CORE} proved to be a crucial step. 
Furthermore, the configuration of the cluster to achieve the greatest possible resource allocation was achieved by setting the following properties: The configuration of the \texttt{maximizeResourceAllocation} parameter to \texttt{true} ensured that the available resources were optimally utilised, particularly in instances where the system was not operating in a fleet mode.

It was necessary to gain an understanding of the specific challenges presented by EMR in relation to memory management. 
By doubling the memory allocations, a significant improvement in performance was achieved, while maintaining cost-effectiveness. 
However, these adjustments often resulted in longer run times compared to DBR. 
This approach highlighted the necessity of striking a delicate balance between cost and performance when utilising EMR.

In order to address the issue of slower operations, particularly those associated with vacuum tasks in Delta Lake, it was necessary to set the parameter \texttt{spark.databricks.delta.vacuum.parallelDelete.enabled} to \texttt{true}. 
This configuration was of paramount importance for optimising maintenance tasks, which could otherwise result in an unnecessarily extended runtime. 
This was automatically handled on Databricks, but required several hours and experimentation until the EMR parameter was identified.

Each of these configuration insights was only realised through continuous refinement and adaptation, demonstrating the necessity of a deeply engaged, experimental approach to managing and optimising a platform like EMR. 
The learning process involved was non-trivial, highlighting the sophisticated understanding required to ensure EMR's operational efficiency and stability. 
This iterative learning and adaptation process is visually represented in the cumulative line chart, which reflects the extensive effort and incremental advancements over time, culminating in a robust, cost-efficient platform setup.

It is also noteworthy that the custom enhancements described in the preceding section, including the Dagster Context Injector, Message Reader, and Cloud Client Innovations, were developed to integrate harmoniously with our existing infrastructure, offering bespoke solutions that generic implementations in the open-source Dagster project may not accommodate. 
The incorporation of specific features into the open-source Dagster project may result in the dilution of their particular characteristics or necessitate extensive modifications that may not be universally applicable or advantageous. 
Consequently, while we endeavour to contribute to the community wherever feasible, some of our modifications will remain open source but separate from the Dagster repository in order to preserve their bespoke functionality. 
Our architectural approach necessitates the construction of data processing environments that are not integrated with Dagster, but rather built upon it. 
This methodology allows us to leverage Dagster's core capabilities while enhancing them with bespoke solutions that align with our strategic goals and technical requirements.

\hypertarget{platform-comparison}{%
\section{Platform Comparison}\label{platform-comparison}}

Table \ref{tab:costTable} provides a comprehensive breakdown of the computational costs associated with processing the same batch of Common Crawl data across EMR and DBR. 
It is evident that utilising DBR significantly increases the overall cost of utilising the platform, particularly for compute-intensive and prolonged operations such as processing the Common Crawl (CC) edges. 
Even for less demanding tasks, the surcharges imposed by DBR represent a significantly higher proportion of the overall expenses, rendering it a costlier option.
However, despite these higher costs, DBR offers the advantage of enhanced performance capabilities, attributed to its sophisticated and optimised data processing frameworks. 
This efficiency potentially justifies the additional financial outlay by contributing to time savings in both operational and developmental phases, which is a critical consideration for projects where speed and performance are prioritised alongside cost.

\begin{landscape}
\begin{table}[]
\begin{tabular}{llllllllll}
\hline
Run & Step        & Platform & Duration & Total Cost & Platform Surcharge & EBS Cost & EC2 Cost & Aggregated Total Cost & Aggregated Total Surcharge \\ \hline
1   & nodes       & EMR      & 0.35     & \$0.40     & \$0.07             & \$0.01   & \$0.32   & \$422.95              & \$90.17                    \\
1   & edges       & EMR      & 9.99     & \$402.54   & \$80.19            & \$13.72  & \$308.63 &                       &                            \\
1   & graph       & DBR      & 0.38     & \$18.30    & \$9.78             & \$0.08   & \$8.44   &                       &                            \\
1   & graph\_aggr & EMR      & 0.27     & \$1.71     & \$0.13             & \$0.02   & \$1.56   &                       &                            \\ \hline
2   & nodes       & DBR      & 0.23     & \$0.50     & \$0.13             & \$0.00   & \$0.37   & \$784.64              & \$252.74                   \\
2   & edges       & DBR      & 5.71     & \$766.17   & \$240.79           & \$22.47  & \$502.91 &                       &                            \\
2   & graph       & DBR      & 0.38     & \$17.04    & \$11.61            & \$0.26   & \$5.17   &                       &                            \\
2   & graph\_aggr & DBR      & 0.11     & \$0.93     & \$0.21             & \$0.00   & \$0.72   &                       &                            \\ \hline
3   & nodes       & EMR      & 0.43     & \$0.42     & \$0.06             & \$0.00   & \$0.36   & \$417.06              & \$83.37                    \\
3   & edges       & EMR      & 10.49    & \$409.03   & \$82.19            & \$13.82  & \$313.02 &                       &                            \\
3   & graph       & EMR      & 0.94     & \$4.71     & \$1.05             & \$0.07   & \$3.59   &                       &                            \\
3   & graph\_aggr & EMR      & 0.23     & \$2.90     & \$0.07             & \$0.00   & \$2.83   &                       &                            \\ \hline
\end{tabular}
\caption{Overview of Computational Costs Across Pipeline Configuration.}
\label{tab:costTable}
\end{table}
\end{landscape}

\begin{figure}[H]
\centering
\includegraphics[width=\textwidth]{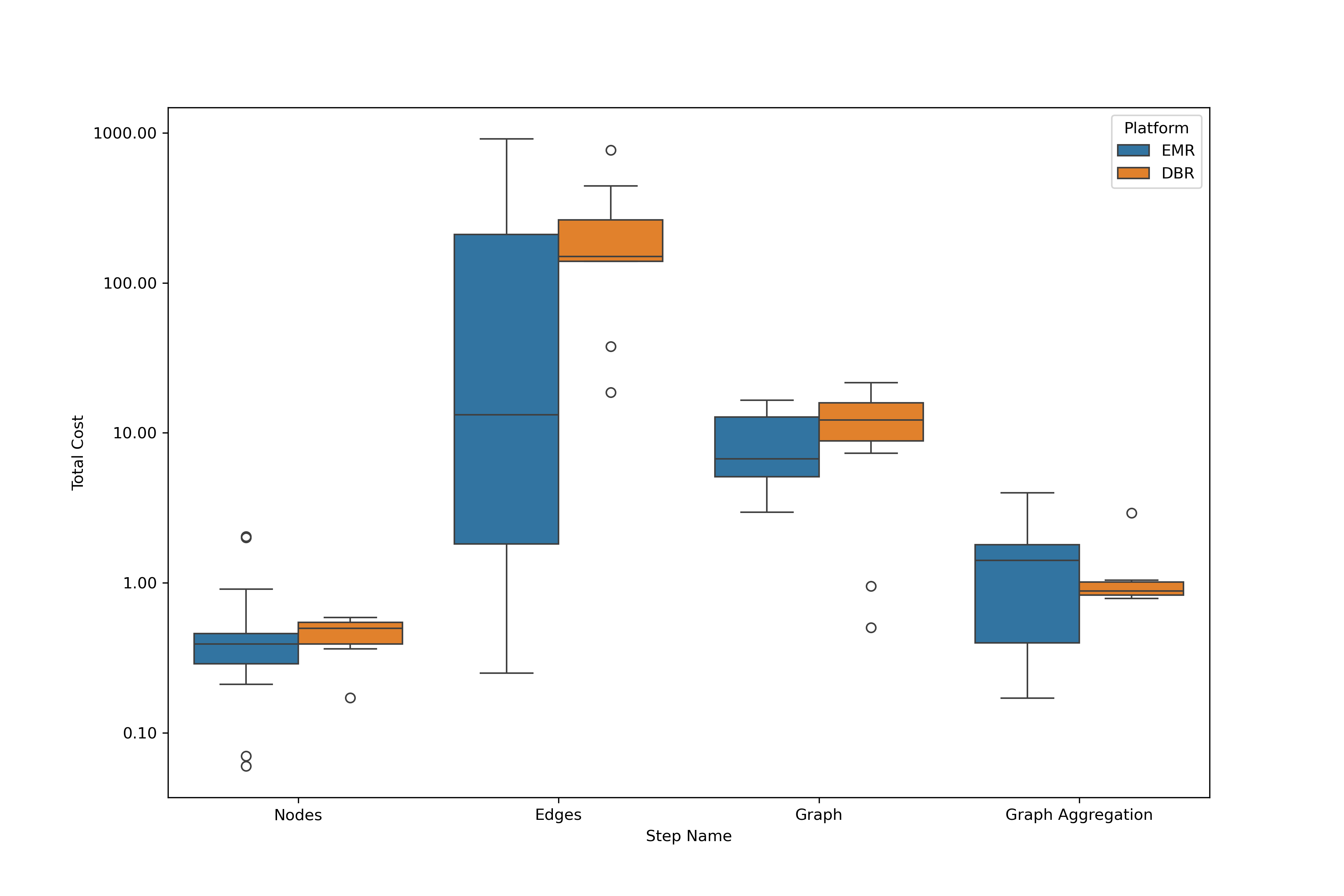}
\caption{Total Cost Production Runs by Asset.\label{fig:costProduction}}
\end{figure}

Figure \ref{fig:costProduction}, unlike the previous table, shows costs for multiple batches of Common Crawl data across various tasks and platforms.
DBR shines in compute-heavy tasks, offering superior performance at a premium. 
Its edge processing capabilities justify higher costs by slashing operational and development time. EMR, on the other hand, proves more economical for simpler operations. 
This cost-effective solution suits budget-conscious projects needing scalable data processing. 
The graph  illustrates these economic trade-offs across various data processing stages, helping institutions and businesses make informed decisions based on their specific needs and financial constraints.

\begin{figure}[H]
\centering
\includegraphics[width=\textwidth]{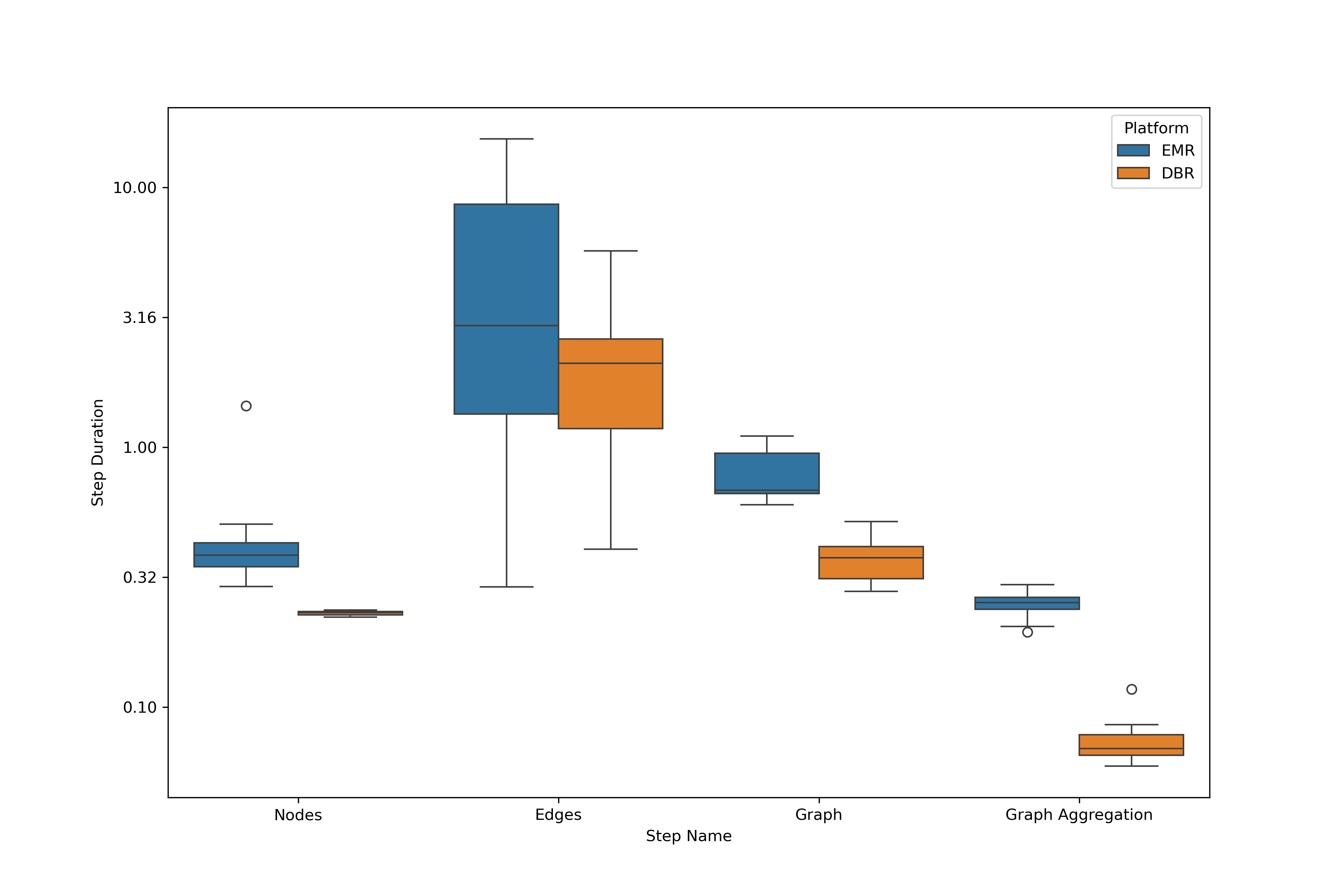}
\caption{Time run distribution on production runs.\label{fig:durationProduction}}
\end{figure}

Figure \ref{fig:durationProduction} offers an empirical comparison of step durations across platforms during different phases of data processing. 
This visualisation demonstrates a notable discrepancy in performance, with DBR exhibiting a consistently reduced processing time. 
This discrepancy in performance can be attributed to DBR' optimised runtime environment, which includes an optimised version of Spark itself as well as a C-based rewrite (Photon Behm et al., 2022) that significantly reduces overhead and execution time. 
Furthermore, DBR offers pre-configured settings, enabling more efficient data processing without the necessity for extensive manual tuning, which is required with EMR. 
This not only underscores the enhanced usability of DBR but also highlights its cost-effectiveness in terms of resource utilisation and time savings.

\hypertarget{acknowledgements}{%
\section{Acknowledgements}\label{acknowledgements}}

This research was supported by \href{https://ascii.ac.at/}{Supply Chain Intelligence Institute Austria (ASCII)}.

\hypertarget{references}{%
\section*{References}\label{references}}
\addcontentsline{toc}{section}{References}

\hypertarget{refs}{}
\leavevmode\hypertarget{ref-Behm}{}%
Behm, A., Palkar, S., Agarwal, U., Armstrong, T., Cashman, D., Dave, A.,
Greenstein, T., Hovsepian, S., Johnson, R., Sai Krishnan, A., \& others.
(2022). Photon: A fast query engine for lakehouse systems.
\emph{Proceedings of the 2022 International Conference on Management of
Data}, 2326--2339.

\leavevmode\hypertarget{ref-dagster}{}%
Dagster. (2018). Dagster \textbar{} cloud-native orchestration of data
pipelines. In \emph{GitHub repository}. GitHub.
\url{https://github.com/dagster-io/dagster}

\leavevmode\hypertarget{ref-Daw}{}%
Daw, N., Bellur, U., \& Kulkarni, P. (2021). Speedo: Fast dispatch and
orchestration of serverless workflows. \emph{Proceedings of the ACM
Symposium on Cloud Computing}, 585--599.
\url{https://doi.org/10.1145/3472883.3486982}

\leavevmode\hypertarget{ref-Heiler}{}%
Heiler, G., \& Picatto, H. (2024). \emph{Cost efficient alternative to
databricks lock-in}.
\href{https://georgheiler.com/2024/06/21/cost-efficient-alternative-to-databricks-lock-in/}{georgheiler.com/2024/05/02/cost-efficient-alternative-to-databricks-lock-in}

\leavevmode\hypertarget{ref-kinne}{}%
Kinne, J., \& Axenbeck, J. (2020). Web mining for innovation ecosystem
mapping: A framework and a large-scale pilot study. \emph{Scientometrics}, \emph{125}(3), 2011--2041.
Springer. \url{https://doi.org/10.1007/s11192-020-03740-9}

\leavevmode\hypertarget{ref-Anil}{}%
Mathew, A., Andrikopoulos, V., Blaauw, F. J., \& Karastoyanova, D.
(2024). Pattern-based serverless data processing pipelines for
function-as-a-service orchestration systems. \emph{Future Generation
Computer Systems}, \emph{154}, 87--100.
https://doi.org/\url{https://doi.org/10.1016/j.future.2023.12.026}

\leavevmode\hypertarget{ref-Zaharia}{}%
Zaharia, M., Xin, R. S., Wendell, P., Das, T., Armbrust, M., Dave, A.,
Meng, X., Rosen, J., Venkataraman, S., Franklin, M. J., Ghodsi, A.,
Gonzalez, J., Shenker, S., \& Stoica, I. (2016). Apache spark: A unified
engine for big data processing. \emph{Commun. ACM}, \emph{59}(11),
56--65. \url{https://doi.org/10.1145/2934664}

\end{document}